\documentclass[aps,pre,preprint,groupedaddress,showpacs]{revtex4-1}
\usepackage{amssymb}

\usepackage[dvips]{graphicx}
\usepackage{textcomp}
\usepackage{amssymb}

\newcommand{\mc}{\multicolumn}

\begin{document}

\title{
\Large\bf A finite size scaling study of lattice models in \\
the three-dimensional Ising  universality class}

\author{Martin Hasenbusch}
\email[]{Martin.Hasenbusch@physik.hu-berlin.de}
\affiliation{
Institut f\"ur Physik, Humboldt-Universit\"at zu Berlin,
Newtonstr. 15, 12489 Berlin, Germany}

\date{\today}

\begin{abstract}
We study the spin-1/2 Ising model and the Blume-Capel model at various 
values of the parameter $D$ on the simple cubic lattice. To this end we 
perform Monte Carlo simulations using a hybrid of the local Metropolis,
the single cluster and the wall cluster algorithm.
Using finite size scaling we determine the value $D^*=0.656(20)$ of the 
parameter $D$, where leading corrections to scaling vanish. We find 
$\omega=0.832(6)$ for the exponent of leading corrections to scaling.
In order to compute accurate estimates of critical exponents we 
construct improved observables that have a small amplitude of the leading 
correction for any model. 
Analyzing data obtained for $D=0.641$ and $0.655$ on lattices of a linear 
size up to $L=360$ we obtain $\nu=0.63002(10)$ and
$\eta=0.03627(10)$. We compare our results with those obtained from previous
Monte Carlo simulations and high temperature series expansions of lattice 
models, by using field theoretic methods and experiments.
\end{abstract}
\maketitle

\section{Introduction}
In the neighborhood of a second order phase transition various quantities 
diverge. For example the correlation length, which characterizes the decay 
of the two-point correlation function, behaves as 
\begin{equation}
\label{xipower}
\xi = f_{\pm} |t|^{-\nu} 
\times \left(1 +  b_{\pm} |t|^{\theta} +  c t
 + d_{\pm} |t|^{\theta'} + e_{\pm}  |t|^{2 \theta}  + ... \right)  \;,
\end{equation}
where $t=(T-T_c)/T_c$ is the reduced temperature, $f_{+}$ and $f_{-}$ are the 
amplitudes in the high and the low temperature phase, respectively, and $\nu$
is the critical exponent of the correlation length. These power laws
are affected by confluent corrections, such as $b_{\pm} |t|^{\theta}$, 
$d_{\pm} |t|^{\theta'}$, $e_{\pm}  |t|^{2 \theta}$, and non-confluent ones 
such as $c t$. 
Critical exponents such as $\nu$ and ratios of amplitudes such as $f_+/f_-$ are
universal. This means that they assume exactly the same value for any system 
within a given universality class. Also correction exponents  such as
$\theta=\omega \nu \approx 0.5$ and ratios of correction amplitudes as 
$b_+/b_-$ are universal.
A universality class is characterized by the dimension of the system, the range
of the interaction and the symmetry of the order parameter. 
The critical exponents $\alpha$ of the specific heat, $\gamma$ of
the magnetic susceptibility, $\eta$ of the two-point correlation function
at the critical point, $\nu$ of the correlation length, $\delta$ of the 
magnetization at the critical temperature as a function of the external field
and $\beta$ of the spontaneous magnetization at a vanishing external field
are related by so called scaling and hyperscaling relations. This allows to
deduce all of them from two independent exponents.
For reviews on critical phenomena and the Renormalization Group (RG) see e.g.
\cite{WiKo,Fisher74,Fisher98,PeVi02}.

Here we are concerned with three dimensions, short range
interactions and a $\mathbb{Z}_2$ symmetry of the order parameter. The
best know model undergoing a phase transition in  this universality class 
is the spin-1/2 Ising model in three dimensions with nearest neighbor
interactions.  Therefore this universality class is called the 
three-dimensional Ising universality class.
This universality class is supposed to be realized in a huge range of 
experimental systems: binary mixtures, uniaxial magnets or micellar 
systems; see e.g. \cite{PeVi02,BaHaLiDu07,SeSh09}. 
Typically the estimates of critical exponents extracted from experimental data
are less accurate than those obtained by using the theoretical methods 
discussed below.  For example, 
recent experimental estimates obtained from turbidity data for a 
methanol-cyclohexane mixture are $\nu=0.632(2)$ and $\eta=0.041(5)$
\cite{LyJa04}. 

Critical exponents and amplitude ratios have been computed by 
various theoretical methods  such as field theoretic methods or
high temperature series expansions and Monte Carlo simulations of lattice
models. 
First let us briefly discuss results obtained by the $\epsilon$-expansion 
\cite{WiFi73}, where the dimension $d$ of the system is given
by $d=4-\epsilon$ and the perturbative expansion in $d=3$ \cite{Pa73}.  
The $\epsilon$-expansion of critical exponents has been computed up to 
$O(\epsilon^5)$ \cite{E5}
while the perturbative expansion in $d=3$ has been computed up to 
seven loops \cite{Guelph}  for the Ising universality class.
Since both expansions are divergent, some kind of resummation is needed
to extract numerical results for critical exponents. 
In the case of the $\epsilon$-expansion  the estimates
reported in the literature are consistent among each other. 
As a representative result we report in table  \ref{compare_field}
the one of ref. \cite{GuZi98}. 
In table  \ref{compare_field} we also give results obtained  
from the perturbative expansion in $d=3$ using different 
resummation techniques. For a more complete compilation see e.g. ref. 
\cite{PeVi02}. In table \ref{compare_field} we give the 
exponents $\nu$, $\eta$ and the correction exponent $\omega$, since 
these are directly computed by using field theoretic methods.
In addition we report the value of $\gamma$ that can be compared with the 
results of the high temperature series expansions reported below.
Typically, the errors reported
for the critical exponents obtained from the perturbative expansion in $d=3$ 
are smaller than those obtained from the $\epsilon$-expansion. 
While the estimates for $\nu$ are all consistent within the quoted errors, 
clear variations can be observed for $\eta$, $\gamma$ and $\omega$.
For a discussion of the different resummation schemes that have been used,
we refer the reader to ref. \cite{PoSu08}.

\begin{table}
\caption{\sl \label{compare_field} Numerical results for the critical 
exponents $\nu$, $\gamma$, $\eta$ and $\omega$ obtained by using field theoretic
methods. The list is by far not exhaustive. We try to give extreme examples;
both concerning the values found as well as the quoted error bar.  In the
case of the $\epsilon$-expansion we have taken the
results that fulfil the boundary condition that for $\epsilon=2$ the
correct 2D Ising results are obtained.
}
\begin{center}
\begin{tabular}{cclllll}
\hline
 \multicolumn{1}{c}{ref}
 & \multicolumn{1}{c}{year}
 & \multicolumn{1}{c}{Method}
 & \multicolumn{1}{c}{$\nu$}
 & \multicolumn{1}{c}{$\gamma$}
 & \multicolumn{1}{c}{$\eta$}
 & \multicolumn{1}{c}{$\omega$} \\
\hline
\cite{GuZi98} & 1998 &$\epsilon$-exp&  0.6305(25)& 1.2380(50) &0.0365(50)&0.814(18)\\
\cite{Nickel} & 1991 &3D exp        &  0.630     & 1.238      &0.0355    &0.845 \\
\cite{Guelph}   & 1991 & 3D exp       &  0.6301(5) & 1.2378(6)  &0.0355(9) &       \\
\cite{GuZi98} & 1998 &3D exp        &  0.6304(13)& 1.2396(13) &0.0335(25)&0.799(11)\\
\cite{Kl99}   & 1999 &3D exp        &  0.6305    & 1.241      &0.0347(1) &0.805   \\
\cite{KlJa01} & 2001 &3D exp        &  0.6303(8) & 1.2403(8)  &0.0335(6) &0.792(3) \\
\cite{PoSu08} & 2008 &3D exp        & 0.6306(5)  & 1.2411(6)  &0.0318(3) &0.782(5) \\
\hline
\end{tabular}
\end{center}
\end{table}

In table \ref{compare_lattice} we summarize recent results obtained from 
lattice models. For an exhaustive summary of previous 
works  see ref. \cite{PeVi02}. The authors of \cite{pisa_series2} have 
analyzed the high temperature series expansion of improved models on the simple 
cubic lattice up to $O(\beta^{25})$, where $\beta=1/k_B T$ is the inverse 
temperature. One of these models is studied here
using Monte Carlo simulations. Improved means that the amplitudes of 
leading corrections to scaling such as $b_{\pm}$ in eq.~(\ref{xipower}) 
vanish. The authors of \cite{BuCo02} have studied
the high temperature series expansion of spin-S Ising models on the 
simple cubic and the body centered cubic lattice
up to $O(\beta^{25})$. Note that in the spin-S Ising model the spin-variable
might assume the values $-S$, $-S+1$,...,$S-1$, $S$. In ref. \cite{BuCo05}
the same authors have studied the $\phi^4$ model on the simple cubic and 
the body centered cubic lattice also up to $O(\beta^{25})$.
These results from high temperature series expansions are
all compatible among each other. Note that these expansions were performed for
lattice models with quite different Hamiltonians. Furthermore, there are 
results for both simple cubic and body centered cubic lattices.
It is highly plausible that corrections to scaling have different amplitudes
in these different models. Therefore the agreement of the results gives us
confidence that there are no undetected systematic errors due to leading,
or in the case of improved models, subleading corrections to scaling. 
The results for the exponents $\nu$, $\gamma$ and $\eta$ obtained from 
the high temperature series expansion are clearly more precise than those
obtained by using field theoretic methods. The results obtained for $\nu$
using field theoretic methods and high temperature series expansions
of lattice models are consistent. In the case of $\gamma$ and $\eta$
some of the results obtained by resumming the perturbative expansion in three 
dimensions can be clearly ruled out by the high temperature series expansion.
Unfortunately, the analysis of the high temperature series expansions
does not provide an accurate estimate for the correction exponent $\omega$.

Lattice models can also be studied by using Monte Carlo simulations. The 
finite size scaling (FSS) approach \cite{Bi81,Barber,Privman} is well suited to
locate the critical temperature and to compute critical exponents.
Typically one simulates the model directly at the critical point. The 
critical exponents are then extracted from the scaling of the observables
with the lattice size. For example, at the critical temperature the 
magnetic susceptibility behaves as 
\begin{equation}
\label{chifi}
\chi =a L^{2-\eta} \times (1 + b L^{-\omega} + c L^{-\omega'} +
                                  d L^{-2 \omega}  + ...) \;+B \;\;,
\end{equation}
where $B$ is an analytic background and $L$ the linear size of a cubic 
lattice with periodic boundary conditions.
An exhaustive summary of previous works  is given in table 5 of 
ref. \cite{PeVi02}. In table \ref{compare_lattice} we only quote  
recent works. In 1999 four finite size scaling studies of lattices models 
in the Ising universality class had been published. The results of these
works are consistent among each other and the accuracy that had been achieved
is similar to that of the field theoretic calculations. I like to mention 
that in \cite{BlShTa99} a special purpose computer for the cluster 
simulation of the Ising model had been used.
In the most recent work \cite{DengBloete}, which provides the most accurate 
estimates so far, 11 different models were studied on lattices up to a linear
site of $L=128$. The results obtained for $\nu$, $\gamma$ and $\eta$ are 
essentially consistent with those obtained from the high temperature 
series expansions. The estimate obtained for $\omega$ is more accurate
than that of the high temperature series expansion and it is 
clearly larger than most of the estimates obtained from the perturbative
expansion in three dimensions.

\begin{table}
\caption{\sl \label{compare_lattice} Numerical results for the critical 
exponents $\nu$, $\gamma$, $\eta$ and $\omega$  obtained by analyzing high
temperature series (HT) and Monte Carlo (MC) simulations of lattice models 
in the Ising universality class. In the case of the Monte Carlo (MC) 
simulations, some of the authors have quoted the statistical and the 
systematical errors of $\nu$ and $\eta$ separately.
The numbers marked by $^*$ are not directly given by the
authors but are computed by using the scaling relation $\gamma = \nu (2-\eta)$.
Note that the error of $\gamma$ is computed naively, assuming that the errors
of $\nu$ and $\eta$  are purely statistical and that the estimates of $\nu$ 
and $\eta$ are uncorrelated.
For an exhaustive summary of previous work see ref. \cite{PeVi02}. 
}
\begin{center}
\begin{tabular}{cccllll}
\hline
 \multicolumn{1}{c}{ref}
 & \multicolumn{1}{c}{year}
 & \multicolumn{1}{c}{Method}
 & \multicolumn{1}{c}{$\nu$}
 & \multicolumn{1}{c}{$\gamma$}
 & \multicolumn{1}{c}{$\eta$}
 & \multicolumn{1}{c}{$\omega$} \\
 \hline
\cite{pisa_series2} & 2002 &HT & 0.63012(16) & 1.2373(2) &  0.03639(15) & 0.825(50)\\
\cite{BuCo02}       & 2002 &HT & 0.6299(2)   & 1.2371(1) &  0.0360(8)$^*$ & --     \\
\cite{BuCo05}       & 2005 &HT & 0.6301(2)   & 1.2373(2) &  0.0363(9)$^*$ & --     \\
\cite{BaFeMMMSPaRL}& 1999 &MC & 0.6294(5)[5]& 1.2353(21)$^*$& 0.0374(6)[6]& 0.87(9) \\
\cite{HaPiVi99} &1999 &MC & 0.6298(2)[3]&1.2365(11)$^*$ & 0.0366(6)[2]& -- \\ 
\cite{Ha99}     &1999 &MC & 0.6296(3)[4]&1.2367(15)$^*$ & 0.0358(4)[5]& 0.845(10) \\
\cite{BlShTa99} &1999 &MC & 0.63032(56) &1.2372(13)$^*$  & 0.0372(10)& 0.82(3) \\
\cite{DengBloete}& 2003 &MC & 0.63020(12) & 1.2372(4)$^*$ & 0.0368(2) & 0.821(5)\\
\hline
\end{tabular}
\end{center}
\end{table}

The purpose of the present work is to corroborate the lattice results
discussed above. To this end we shall simulate lattices that are 
considerably larger than those of ref. \cite{DengBloete}. Furthermore 
we shall use improved observables that have been applied in ref. 
\cite{ourdilute} to study Ising models with quenched dilution.  Here,
improved means that the amplitude of the leading correction vanishes for 
any model. Since these observables are constructed numerically, in practice
some residual amplitude remains. Using these improved 
observables in the study of improved models, leading corrections are 
highly suppressed, allowing us to ignore them in the finite size scaling 
analysis.

Accurate numerical estimates of critical exponents might serve as  benchmark 
for future experiments, see e.g. \cite{BaHaLiDu07}, the analysis of the 
perturbative expansion in three dimensions, as discussed above, or new 
theoretical approaches such as new ideas in the so called exact 
renormalization group \cite{exact} or the Kallen-Lehmann approach 
\cite{Kallen}.

The outline of the paper is the following.  First we define the model and 
the observables that are studied. Then we discuss the Monte Carlo algorithm
that has been used. We give the details of our numerical study.
We estimate the fixed point values of
the phenomenological couplings and the inverse transition temperatures.
We give a numerical estimate of the correction exponent $\omega$ and obtain
a new estimate of $D^*$, the value of the parameter where leading corrections
to scaling vanish. Next we construct various improved observables. Based on this
we compute estimates for the critical exponents $\nu$ and $\eta$.

\section{The model}

The spin-1/2 Ising model is characterized by the reduced Hamiltonian
\begin{equation}
\label{Isingaction}
H = -\beta \sum_{<xy>} s_x s_y
  - h \sum_x s_x \;\;  ,
\end{equation}
where the spin might assume the values $s_x \in \{-1, 1 \}$. $x=(x_0,x_1,x_2)$
denotes a site of the simple cubic lattice, where $x_i \in \{0,1,2,...,L_i-1\}$.
$<xy>$ denotes a pair of nearest neighbors on the lattice. We employ periodic
boundary conditions in all directions of the lattice.
Throughout we shall consider $L_0=L_1=L_2=L$ and a vanishing 
external field $h=0$. The partition function is given by
\begin{equation}
 Z = \sum_{\{s_x\}} \exp(-H) \;\;  ,
\end{equation}
where $\sum_{\{s_x\}}$ denotes the sum over all configurations.

The Blume-Capel model is characterized by the reduced Hamiltonian
\begin{equation}
\label{Blumeaction}
H = -\beta \sum_{<xy>} s_x s_y
  + D \sum_x s_x^2  - h \sum_x s_x \;\;  ,
\end{equation}
where now the spin might assume the values $s_x \in \{-1,0,1 \}$.
In the limit $D \rightarrow - \infty$  the ``state"  $s=0$ is completely
suppressed, compared with $s=\pm 1$, and therefore the spin-1/2 Ising model 
is recovered.
In  $d\ge 2$  dimensions the model undergoes a continuous phase transition
for $-\infty \le  D   < D_{tri} $ at a $\beta_c$ that depends on $D$. 
For $D > D_{tri}$ the model undergoes a first order phase transition.
Refs. \cite{des,HeBlo98,DeBl04}  give for the three-dimensional simple cubic 
lattice
$D_{tri} \approx 2.006$,  $D_{tri}\approx 2.05$ and $D_{tri} =2.0313(4)$,
respectively.

Numerically it has been shown that on the line of second order phase 
transitions there is a point, where leading corrections to scaling vanish.
In the following we shall call the model at this point ``improved model''.
In ref. \cite{myhabil} we find $D^*=0.641(8)$. One should note that no
effort was made to estimate the systematical error due to subleading 
corrections to scaling.
The authors of \cite{BlLuHe95,DengBloete} have simulated the model at 
$D=\ln 2 = 0.693147...$ . At this value of $D$ corrections to scaling are still 
small compared with the spin-1/2 Ising model. At $D=\ln 2$ the 
Blume-Capel model can be mapped into a spin-1/2 Ising model with twice 
the number of sites. This model can be simulated with a cluster algorithm
without additional local updates as it is the case for general values of $D$. 

\section{The observables}
The energy of a given spin configuration is defined as 
\begin{equation}
\label{energy}
 E=  \sum_{<xy>}  s_x  s_y \;\;.
\end{equation}
This definition is convenient for our purpose. One should note however that
it deviates from the standard textbook definition.
The magnetic susceptibility $\chi$ and the second moment correlation length 
$\xi_{2nd}$ are defined as
\begin{equation}
\chi  \equiv  \frac{1}{V}
\left\langle \Big(\sum_x s_x \Big)^2 \right\rangle \;\;,
\end{equation}
where $V=L^3$ and
\begin{equation}
\xi_{2nd}  \equiv  \sqrt{\frac{\chi/F-1}{4 \sin^2 \pi/L}} \;\;,
\end{equation}
where
\begin{equation}
F  \equiv  \frac{1}{V} \left \langle
\Big|\sum_x \exp\left(i \frac{2 \pi x_k}{L} \right)
        s_x \Big|^2
\right \rangle
\end{equation}
is the Fourier transform of the correlation function at the lowest
non-zero momentum. In our simulations, we have measured $F$ for the three
directions $k=0,1,2$ and have averaged these three results.

In addition to elementary quantities like the energy, the
magnetization, the specific heat or the magnetic susceptibility, we
compute a number of so-called phenomenological couplings, that means
quantities that, in the critical limit, are invariant under RG
transformations. 
We consider the Binder parameter $U_4$ and its sixth-order generalization
$U_6$, defined as
\begin{equation}
U_{2j} \equiv \frac{\langle m^{2j}\rangle}{\langle m^2\rangle^j},
\end{equation}
where $m = \frac{1}{V} \, \sum_x s_x$ is the
magnetization of a given spin configuration.  We also consider the ratio
$R_Z\equiv Z_a/Z_p$ of
the partition function $Z_a$ of a system with anti-periodic boundary
conditions in one of the three directions and the partition function
$Z_p$ of a system with periodic boundary conditions in all directions.
Anti-periodic boundary conditions in the zero direction are obtained
by replacing $s_x  s_y$ by $-s_x  s_y$ in the
Hamiltonian for links $\left<xy\right>$ that connect the boundaries,
i.e., for $x=(L-1,x_1,x_2)$ and $y=(0,x_1,x_2)$. The ratio $Z_a/Z_p$ 
can be efficiently evaluated using the boundary flip algorithm \cite{BF}. 
Here we use a modified  version of the boundary flip algorithm as
discussed in appendix A 2 of ref. \cite{ourXY}. 
In the following we shall refer to the RG-invariant 
quantities $U_{2j}$, $R_Z\equiv Z_a/Z_p$ and $R_\xi\equiv \xi_{2nd}/L$
using the symbol $R$.

In our analysis we need the observables as a function of $\beta$ in 
some neighborhood of the simulation point.  To this end we have 
computed the coefficients of the Taylor expansion of the observables 
up to the third order. 
For example the first derivative of the expectation value 
$\langle A \rangle$ of an observable $A$ is given by
\begin{equation}
\frac{\partial \langle A \rangle}{\partial \beta} = \langle A E \rangle
- \langle A \rangle \langle E \rangle \;\;.
\end{equation}

\section{The simulation algorithm}

Analogous to \cite{BrTa}, we have simulated the Blume-Capel model
using a hybrid of local updates and  cluster updates. The cluster algorithm 
only changes the sign of spins. Therefore, in order to get an ergodic
algorithm for the Blume-Capel model with finite $D$,
local Metropolis updates are used that also can change the modulus
$|s_x|$ of the spins. Following \cite{PlFeLa02} even in the case of the 
spin-1/2 Ising model such a hybrid of local and cluster updates is 
superior to the cluster algorithm alone. The authors of  \cite{PlFeLa02}
also found that such a hybrid algorithm is much less susceptible to 
systematic errors caused by the imperfection of pseudo-random numbers 
than a pure cluster algorithm.
Here we have used a hybrid of local Metropolis updates that are
implemented by using the multispin coding technique \cite{multispin},  
single cluster updates \cite{Wolff} and wall cluster updates \cite{wall}.
In the single cluster update, the cluster that includes a randomly 
chosen site is flipped. In contrast, in the wall cluster update all
clusters that include sites that are part of a given plane (the "wall") of the 
lattice are flipped.

Motivated by the multispin coding implementation of the local update 
we have simulated $N_{bit} = 64$ copies of the system  in parallel. 
In the first stage of our study, we have used a single random number sequence
for the local Metropolis updates of these $N_{bit}$ systems. This leads
to some degradation of the performance. To diminish this problem, we have used
a modified sequence of the pseudo random numbers in the second stage of our 
study. Details are given below. In the case of the cluster updates we could
not make use of the multispin coding technique. Therefore we have updated 
the systems one by one, using different random number sequences for each of the 
systems.  

Let us discuss the implementation of the local Metropolis algorithm 
in more detail.
We have implemented the spin $s_{x} \in \{-1,0,1 \}$ using two bits. To 
this end we write $s_{x} = \sigma_x \tau_x$, where $\sigma_x \in \{-1,1\}$
and $\tau_x \in \{0,1\}$.  In terms of these new variables the partition 
function becomes 
\begin{equation}
Z = C  \sum_{\{\sigma_x\}} \sum_{\{\tau_x\}} 
   \exp\left( \beta \sum_{<xy>} 
  \sigma_x \sigma_y  \tau_x \tau_y- \tilde D \sum_x \tau_x\right) \;\;,
\end{equation}
where $\tilde D = D-\ln2$. Note that subtracting $\ln 2$ corrects 
for the double counting of the $s_x =0$ state.

In our local updating scheme we performed consecutive updates of 
$\sigma_x$ and $\tau_x$. In the first step, the proposal is given by 
$\sigma_x'=-\sigma_x$.
It is accepted with the standard Metropolis acceptance probability
\begin{equation}
\label{firstacc}
 P_{acc}  = 
\mbox{min}\left[1,\exp\left(-2 \beta \sigma_x \tau_x \sum_{y.nn.x} 
\sigma_y \tau_y \right) \right]  \;\;,
\end{equation}
where $y.nn.x$ means that $y$ is a nearest neighbor of $x$.
In the second step, the proposal is given by $\tau_x'=1-\tau_x$.
A natural choice for the acceptance is
\begin{equation}
P_{acc}  =
\mbox{min}\left[1, 
\exp\left((2 \tau_x-1) \left [-\beta \sigma_x  \sum_{y.nn.x} \sigma_y \tau_y 
   + \tilde D \right] \right) \right] \;\;.
\end{equation}
Instead, for technical reasons we have implemented a two stage acceptance step.
For $D = 0.641$ and $D = 0.655$, where $|\tilde D|$ is small, we have chosen
\begin{equation}
\label{acc1}
P_{acc,1}  =
\mbox{min}\left[1,
\exp\left( \beta (1-2 \tau_x) \sigma_x  \sum_{y.nn.x} \sigma_y \tau_y
   \right) \right] \;
\end{equation}
and
\begin{equation}
\label{acc2}
P_{acc,2}  =
\mbox{min}\left[1,
\exp\left((2 \tau_x-1)  \tilde D \right) \right] \;\;.
\end{equation}
Detailed balance can be easily proven by going through the four cases which are
given by positive or negative arguments of the exponential function in 
Eqs.~(\ref{acc1},\ref{acc2}). 
We take two uncorrelated random numbers $r_1$ and $r_2$ from a uniform 
distribution in $[0,1]$. If both $P_{acc,1} \ge r_1$ and $P_{acc,2} \ge r_2$
the proposal is accepted. 

In the first stage of our study, we have used the same random number 
sequence for all $N_{bit}=64$ systems that we have simulated in parallel. 
In the 
second stage of the study, we have used a modified sequence of the 
random numbers for the acceptance step~(\ref{firstacc}). We have used a
64 bit integer random number in addition to the random number $r$ that
is uniformly distributed in $[0,1]$. If the $i^{th}$ bit of this integer
random number is $1$ we take $r$ itself for the acceptance step of the $i^{th}$ 
system. Otherwise, if the bit is $0$, we take $1-r$ instead. This 
modification considerably reduces the correlation  
among the  $N_{bit}=64$ systems that are simulated in parallel.

We have compared the performance of this local update with that of a 
local heat bath using a standard implementation. 
To this end, we have simulated a $16^3$ lattice at $\beta=0.3877218$, which 
is close to $\beta_c$ as we shall see below.
The integrated autocorrelation times in units of sweeps
of the energy density and the 
magnetic susceptibility are by a factor of about 1.3 larger for the 
Metropolis update discussed here than for the heat bath update.
One sweep over $N_{bit}=64$ systems in parallel using the multispin coding 
technique takes about 4 times as much CPU-time as one sweep over a single
system using the standard implementation of the heat-bath update.
In order to compare the efficiency of the two local updates, we have
computed the statistical error of the energy and the magnetic susceptibility, 
taking into account the possible correlation among the $N_{bit}=64$ systems
in the case of the multispin coding implementation. To this end 
we have performed a Jackknife analysis, where we have first averaged 
the measurements of the $N_{bit}=64$ systems at a given iteration of the 
Monte Carlo simulation.
Taking the inverse of the statistical error squared times the CPU time needed
as measure of the efficiency, we find a performance gain
of a factor of about 10 of the multispin coding implementation of the
local Metropolis update compared with the standard implementation of the 
heat bath update.

During the simulation, local Metropolis sweeps, single cluster and 
wall cluster updates are performed in a certain sequence. 
In the following we denote an elementary building  block of the 
sequence by cycle.
In the case of 
our most recent simulations ($D=0.655$) such a cycle is composed of
\begin{itemize}
\item
4 $\times$ (2 Metropolis sweeps followed by  $L/16$ single cluster updates)
\item
3 Metropolis sweeps
\item
one wall-cluster update
\end{itemize}
In the case of the wall-cluster update we  chose the wall to be perpendicular 
to the 0, 1 and 2-axis in three subsequent cycles. The position of the wall
along the axis is chosen randomly each time. The parameters of the cycle 
are chosen such that roughly the same amount of CPU-time is spent in 
each of its three components.

In order to study the performance of the algorithm, we have  performed
preliminary simulations for $D=0.655$, where we have determined the
autocorrelation function $\rho(t)$ of the magnetic susceptibility and the energy
density. The statistics of these runs is $300000$ update-cycles for the
lattice sizes $L= 16, 32, 64$, and $128$ and $82000$ update-cycles for $L=256$
at $\beta=0.3877218$, which is close to our final estimate of $\beta_c$. 
The integrated autocorrelation time is given by
\begin{equation}
\tau = \frac{1}{2} + \sum_{t=1}^{t_{max}} \rho(t) \;\;,
\end{equation}
where we have chosen $t_{max} = 6 \tau$, selfconsistently.
Fitting our results for 
integrated autocorrelation times in units of update-cycles we get
\begin{equation}
\label{tau1}
 \tau_{\chi}  = 0.70(4) \times L^{0.34(1)} 
\end{equation}
for the magnetic susceptibility and 
\begin{equation}
\label{tau2}
 \tau_{E} = 0.47(2) \times  L^{0.42(1)}
\end{equation}
for the energy density. This means that the autocorrelation times are only a 
few cycles, even for our largest lattices.

We have estimated the statistical errors of the observables using the
Jackknife method. As input of this analysis we have taken data that are
already averaged over the $N_{bit} = 64$ systems that are simulated in
parallel and might be correlated by the use of a common sequence of
random numbers during the Metropolis updates. Therefore the possible
correlation among these $N_{bit} = 64$ systems does not affect the correctness
of the estimate of the statistical errors.

To figure out  how much this correlation does affect the efficiency of the 
algorithm, we have computed the statistical error, taking only one
system, and for comparison, averaging over all  $N_{bit}$ systems.  
If the simulations 
were independent, the square of the ratio of these errors, denoted by $R^2$
in the following, would be 
equal to $N_{bit}$.  In fact we see some performance loss due to 
the use of a common random number sequence. For $L=16$ we get 
for the energy density $R^2 \approx 28.6$ and for the magnetic susceptibility
$R^2 \approx 36.2$. Fortunately these numbers increase with increasing 
lattice size. For $L=256$ we get for the energy density $R^2 \approx 42.2$ 
and $R^2 \approx 48.8$ for the  magnetic susceptibility. 

In order to give an accurate result for the performance gain that is
achieved by using our particular multispin coding implementation
of the local update, one would have to tune the parameters of the 
update cycle for both types of the local update.  For lack of time 
this could not be done. Since the local update is only one of the 
three components of the complete update cycle, likely the gain is 
moderate, certainly less than a factor of two.

\subsection{The simulations: CPU time and statistics}

In a first stage of the study we have simulated the 
spin-1/2 Ising model and the Blume-Capel model at $D=0.641$, $\ln 2$, 
$1.15$ and $1.5$. Note that $D=0.641$ is the estimate of ref. 
\cite{myhabil} for $D^*$ and $D=\ln 2$ has been simulated before by 
the authors of refs. \cite{BlLuHe95,DengBloete}. At $D=1.15$ the amplitude 
of leading corrections to scaling has about the same magnitude as 
for the spin-1/2 Ising model but opposite sign. A preliminary analysis
of these data resulted in $D^* \approx 0.655$.  Therefore we have simulated
at $D=0.655$  in a second stage of our study. 

We have simulated lattices of a linear size $L$ up to 
$L_{max}=96$, $200$, $360$, 
$300$, $64$ and $48$ for the spin-1/2 Ising model and the Blume-Capel 
model at $D=0.641$, $0.655$, $\ln 2$, $1.15$ and $1.5$, respectively. 
In table \ref{Statistics} we have summarized in detail the lattice 
sizes that we have simulated and the statistics of these simulations.

\begin{table}
\caption{\sl \label{Statistics} We give the number of 
update-cycles divided by $64 \times 15000$ as a function of the 
lattice size and the value of the parameter $D$. 
For a discussion see the text.
}
\begin{center}
\begin{tabular}{lrrrrrr}
\hline
\phantom{0}$L$ \phantom{m} &  Ising  & 0.641  & 0.655 & $\ln 2$ & 1.15 & 1.5 \\
\hline
\phantom{0}10 & 10000 & 10000  &   4005  &         & 10000 &       \\
\phantom{0}11 &       &        &   4005  &         &       &       \\
\phantom{0}12 &  9593 & 20000  &   4000  & 4083    & 10000 &  1000 \\
\phantom{0}13 &       &        &   4005  &         &       &       \\
\phantom{0}14 & 11747 & 10000  &   4005  & 3003    & 10000 &       \\
\phantom{0}15 &       &        &   3994  &         &       &       \\
\phantom{0}16 & 9740  & 20200  &   3999  & 2371   & 9917  &  1000 \\
\phantom{0}17 &       &        &   4000  &         &       &       \\
\phantom{0}18 & 11524 & 10000  &   3993  &  1807   &11102  &       \\
\phantom{0}20 & 6959  & 12000  &   3995  &  1828   &7208   &       \\
\phantom{0}22 & 11320 &        &   4003  &  1813   & 12328 &       \\
\phantom{0}24 & 12000 & 10971  &   4000  &  2471   & 13239 &  1000 \\
\phantom{0}28 & 4374  & 7024   &   3999  &  2920   & 5420  &       \\
\phantom{0}32 & 5091  & 2291   &   3011  &  2533   & 4582  &  692  \\
\phantom{0}36 & 3951  & 3362   &         &         & 3444  &       \\
\phantom{0}40 & 1658  & 2349   &   1657  &  1466   & 1474  &       \\
\phantom{0}48 & 1890  & 2202   &         &         & 1502  &  206  \\
\phantom{0}50 &       &        &  968    &   624   &       &       \\
\phantom{0}56 &  629  & 824    &         &         &  688  &       \\
\phantom{0}64 &  719  & 697    &  286    &         &  485  &       \\
\phantom{0}70 &       &        &  894    &        &        &       \\
\phantom{0}72 &       & 848    &         &        &        &       \\
\phantom{0}80 &       & 1053   &         &        &       &       \\
\phantom{0}96 &  273  &        &         &       &        &       \\
100&       &753     &  435    &       &        &       \\
128&       &        &  136    &       &       &       \\
150&       & 319    &  179    &       &       &       \\
200&       & 149    &  106    &      &        &       \\
250&       &        &  118    &  58  &        &       \\
300&       &        &   62    &  14  &       &      \\
360&       &        &   11    &      &       &     \\
\hline
\end{tabular}
\end{center}
\end{table}
In total we have spent the equivalent of $3.5$, $9$, $16$, $3$, $3$, $0.1$ 
CPU years on a single core of a  Quad-Core AMD Opteron(tm) Processor 2378
running at 2.4 GHz for the spin-1/2 Ising model and the
Blume-Capel model at $D=0.641$, $0.655$, $\ln 2$, $1.15$ and $1.5$, 
respectively.

As random number generator we have used the  SIMD-oriented Fast
Mersenne Twister algorithm \cite{twister}. As a check we have repeated 
our simulations at $D=0.655$ using the WELL Random number generator \cite{well}
with about one third of the statistics reported in table \ref{Statistics}.
In particular we have used the program ``WELL44497a.c'' provided by the authors.
We found that the estimates of individual observables are
consistent. We have also repeated part of the finite size scaling analysis 
using these data.  For given ans\"atze we found consistent, 
even though less precise results for the critical exponents. 
One should note that the statistical error
of the fit parameters is often much smaller than the final error that 
also includes systematical errors due to subleading corrections.
The following analysis is only based on the simulations using the 
Mersenne Twister algorithm \cite{twister}.

\section{$\beta_c$ and the fixed point values of phenomenological 
couplings}
In a first step of the analysis we have studied the finite 
size scaling behavior of the phenomenological couplings at 
$D=0.655$, since here we have accumulated  the best 
statistics and secondly, as we shall see below, this value of 
$D$ is closest to $D^*$ among the values that we have simulated.

At the critical point a phenomenological coupling behaves as 
\begin{equation}
\label{Rcorrection}
R(L,\beta_c) = R^* + a L^{-\omega} + b L^{-\omega'} + c L^{-2 \omega} + ...\;,
\end{equation}
where $\omega \approx 0.8$ as discussed in the introduction. Below we 
shall find $\omega=0.832(6)$. The subleading corrections exponent is
$\omega' =1.67(11)$ \cite{NewmanRiedel}. Furthermore, there should be 
corrections with $\omega'' \approx 2$ due to the breaking of the rotational 
symmetry by the lattice \cite{pisa97} or due to the analytic background of the 
magnetic susceptibility.  Motivated by eq.~(\ref{Rcorrection}), 
we have fitted our data with three different ans\"atze
\begin{eqnarray}
  \label{xxxx1}
 R(L,\beta_c) &=& R^* \\
  \label{xxxx2}
 R(L,\beta_c) &=& R^* + a L^{-\epsilon_1}  \\
  \label{xxxx3}
 R(L,\beta_c) &=& R^* + a L^{-\epsilon_1} + b L^{-\epsilon_2}  \;\;,
\end{eqnarray}
where
we have used in eq.~(\ref{xxxx2}) the choices $\epsilon_1=0.83$, 
$\epsilon_1=1.6$ or $\epsilon_1=2$ and in eq.~(\ref{xxxx3})
$\epsilon_1=0.83$ and $\epsilon_2=1.6$ or $\epsilon_2=2$. Here and in the 
following ans\"atze, we denote a correction exponent with a fixed value
by $\epsilon$. Instead, if it is a free parameter we shall denote it, as usual,
by $\omega$.
Here we
need the phenomenological couplings $R$ as a function of the inverse 
temperature. To this end we have used the Taylor-expansion around 
the value $\beta_s$ of the inverse temperature that we have used in the 
simulation. We have checked that the result for $\beta_c$ and $\beta_s$
are sufficiently close to avoid significant truncation effects.
This way, for example eq.~(\ref{xxxx1}) becomes
\begin{equation}
 R(L,\beta_s) = R^* - c_1(L,\beta_s) (\beta_c-\beta_s) 
                    - \frac{c_2(L,\beta_s)}{2!} (\beta_c-\beta_s)^2
                    - \frac{c_3(L,\beta_s)}{3!} (\beta_c-\beta_s)^3 \;\;,
\nonumber
\end{equation}
where $R^*$ and $\beta_c$ are the two parameters of the fit.
Since we have chosen $\beta_s$ as a good approximation of $\beta_c$, 
we could ignore the relatively small statistical error of the Taylor
coefficients $c_1$, $c_2$ and $c_3$, which simplifies the fit.

As an example let us discuss the results obtained for $Z_a/Z_p$ in more 
detail. A selection of our results is given in table \ref{ZAZPS}. 
We have fitted the data for all linear lattice sizes $L$ that are larger than
or equal to a certain $L_{min}$. 
Starting from the $L_{min}$ given in column 5 of table \ref{ZAZPS} the 
$\chi^2/$d.o.f. is close to one. 

\begin{table}
\caption{\sl \label{ZAZPS} Fitting the data for $Z_a/Z_p$ obtained 
at $D=0.655$ with the ans\"atze~(\ref{xxxx1},\ref{xxxx2},\ref{xxxx3}).
$L_{min}$ is the minimal lattice size that is included into the fit.
For a discussion see the text.
}
\begin{center}
\begin{tabular}{ccccllc}
\hline
\mc{1}{c}{ansatz} &  \mc{1}{c}{$\epsilon_1$} &
\mc{1}{c}{$\epsilon_2$} & \mc{1}{c}{$L_{min}$} & \mc{1}{c}{$\beta_c$} &
\mc{1}{c}{$(Z_a/Z_p)^*$} & \mc{1}{c}{$\chi^2/$d.o.f.} \\
\hline
\ref{xxxx1}& -   &  -  &  32 & 0.387721745(10) & 0.542489(14) &
                                                       \phantom{0}9.5/10 \\
\ref{xxxx2}&0.83 &  -  &  18 & 0.387721730(12) & 0.542589(33) & 12.4/14 \\
\ref{xxxx2}&1.6  &  -  &  12 & 0.387721729(10) & 0.542558(12) & 24.0/20 \\
\ref{xxxx2}&2    &  -  &  10 & 0.387721734(10) & 0.542532(8)  & 27.4/22 \\
\ref{xxxx3}&0.83 & 1.6 &  10 & 0.387721746(12) & 0.542448(46) & 27.6/21 \\
\ref{xxxx3}&0.83 &  2  &  10 & 0.387721740(12) & 0.542502(38) & 26.8/21 \\
\hline
\end{tabular}
\end{center}

\end{table}

Taking into account the variation of the results over the different 
ans\"atze we arrive at our final estimate $\beta_c=0.38772174(2)$ and 
$(Z_a/Z_p)^*=0.5425(1)$.  We performed a similar analysis for 
$\xi_{2nd}/L$, $U_4$ and $U_6$.  Our final results are summarized in 
table \ref{betac}.  We find that the estimates of $\beta_c$ obtained 
from different phenomenological couplings are consistent within error bars.
We take the average
\begin{equation}
\label{bc0.655}
\beta_c = 0.387721735(25)
\end{equation}
as our final estimate of the inverse critical temperature. The error bar 
is chosen such that it covers all results given in table \ref{betac}, 
including their error bars.

\begin{table}
\caption{\sl \label{betac}  Results for the inverse critical
temperature $\beta_c$ at $D=0.655$ obtained from the FSS study of 
various phenomenological couplings.
In addition we give the fixed point values $R^*$ of these quantities.
For a discussion see the text.
}
\begin{center}
\begin{tabular}{cllll}
\hline
         &$Z_a/Z_p$   & $\xi_{2nd}/L$ &   $U_4$     &  $U_6$    \\
\hline
$\beta_c$& 0.38772174(2)&0.38772174(2)&0.38772173(2)&0.38772173(2)\\
$R^*$    &0.5425(1)    &0.6431(1)   &1.6036(1)    &3.1053(5)   \\
\hline
\end{tabular}
\end{center}
\end{table}
Our result for $U_4^*$ is about 3 times the combined error smaller 
than  $U_4^*=1/0.62341(4)= 1.60408(10)$ given in \cite{DengBloete}.
We regard our result as more reliable, since we have simulated 
larger lattices and have carefully estimated systematic errors due to
subleading corrections that are not included into the fit.

In the case of the other models we also determined $\beta_c$  by 
fitting with the ans\"atze~(\ref{xxxx1},\ref{xxxx2},\ref{xxxx3}).   Here 
however we have used the results for $(Z_a/Z_p)^*$, $(\xi_{2nd}/L)^*$, 
$U_4^*$ and $U_6^*$ given in table \ref{betac} as input. The final results
obtained this way are summarized in table \ref{betacc}.  For completeness
we have included the results for $D=0.655$ given in eq.~(\ref{bc0.655}). 

\begin{table}
\caption{\sl \label{betacc}  Estimates of the inverse critical
temperature $\beta_c$ of the spin-1/2 Ising model and the Blume-Capel model
at various values of $D$.  For a discussion see the text.
}
\begin{center}
\begin{tabular}{ll}
\hline
  $D$                   &  $\beta_c$    \\
\hline
  Ising                 & 0.22165463(8) \\
  0.641                 & 0.38567122(5) \\
  0.655                 & 0.387721735(25) \\
  $\ln 2 =0.69314718...$& 0.39342239(8) \\ 
  1.15                  & 0.4756110(2)  \\
  1.5                   & 0.5575303(10) \\
\hline                                                                          \end{tabular}                                                                   \end{center}                                                                    \end{table}                  

Our result for $\beta_c$ of the spin-1/2 Ising model is fully consistent 
with $\beta_c=0.22165455(3)$ given in \cite{DengBloete}. For a summary of 
previous results for $\beta_c$ of the spin-1/2 Ising model
we refer the reader to table 1 of \cite{myhabil}.
Our result for $\beta_c$ at $D=\ln 2$ is by $1.5$ times the combined error
larger than $\beta_c=0.39342225(5)$ given in  \cite{DengBloete}.

\section{The correction exponent $\omega$ and the improved model}
In this section we study the cumulants $U_4$ and $U_6$ at a fixed value 
of $Z_a/Z_p$ or $\xi_{2nd}/L$. 
To this end one determines the inverse temperature 
$\beta_f(L)$   defined by
\begin{equation}
\label{definebetaf}
 R_1(L,\beta_f(L)) = R_{1,f} \;\;,
\end{equation}
where $R_1$ is either $Z_a/Z_p$ or $\xi_{2nd}/L$ and $R_{f,1}$ the required 
value. As $R_{f,1}$ we take the fixed point values of $Z_a/Z_p$ and 
$\xi_{2nd}/L$ obtained above. We define
\begin{equation}
 \bar{R}_2(L) \equiv R_2(L,\beta_f(L)) \;\;,
\end{equation}
where $R_2$ is, in our case, either $U_4$ or $U_6$. In the following we shall 
denote $\bar{R}_2$ by $R_2$ at $R_1=R_{1,f}$. In practice we have
done these calculations using the Taylor-expansion of $R_1$ and $R_2$ around 
the value $\beta_s$ that we have used in the simulation up to third order. 
We have checked carefully that $\beta_s$ and $\beta_f$ are sufficiently 
close to avoid significant truncation errors.

One finds, 
see e.g. section III of \cite{ourXY} 
\begin{equation}
\label{barcorrections}
 \bar{R}(D,L)   = \bar{R}^* 
    + a(D) L^{-\omega} + b(D) L^{-\omega'} + ... \;
                           + c \; a^2(D) L^{-2 \omega} + ... \;\;,
\end{equation}
where we should note that the correction amplitudes depend on the parameter
$D$ of our model. The improved model is characterized by a vanishing 
amplitude of leading corrections to scaling. Hence  $D^*$ is given by the 
zero of $a(D)$. 
We have analyzed the data of 5 different models in combined fits:  
The Ising model
and the Blume-Capel model at $D=0.641$, $D=0.655$, $D=\ln 2$ and $D=1.15$. 
To this end we have employed various ans\"atze that are derived from
eq.~(\ref{barcorrections}):
\begin{eqnarray}
\label{barfit1}
 \bar{R}(D,L) &=& \bar{R}^*  +  a(D)  L^{-\omega}  \\
\label{barfit2}
 \bar{R}(D,L) &=& \bar{R}^*  +  a(D)  L^{-\omega}  
                             + c \; a^2(D) L^{-2 \omega}  \\
\label{barfit2x}
 \bar{R}(D,L) &=& \bar{R}^*  +  a(D)  L^{-\omega}
                             + c \; a^2(D) L^{-2 \omega}  
                             + b L^{-\epsilon} \\
\label{barfit3}
 \bar{R}(D,L) &=& \bar{R}^*  +  a(D)  L^{-\omega} 
                             + c \; a^2(D) L^{-2 \omega}  
                             + d \; a^3(D) L^{-3 \omega}  \\
\label{barfit3x}
\bar{R}(D,L) &=& \bar{R}^*  +  a(D)  L^{-\omega}
                             + c \; a^2(D) L^{-2 \omega}
                             + d \; a^3(D) L^{-3 \omega}
                             + b L^{-\epsilon}  \;\;.
\end{eqnarray}
In the ansatz~(\ref{barfit1}) the free parameters of the fit are 
$\bar{R}^*$, $a(\mbox{Ising})$, $a(0.641)$,
 $a(0.655)$,
$a(\ln 2 )$, $a(1.15)$  and the correction exponent 
$\omega$. In the ansatz~(\ref{barfit2}) we have in addition the parameter
$c$. In the ansatz~(\ref{barfit2x}) we have added the term 
$b L^{-\epsilon}$ to take subleading corrections into account. 
Here we make the approximation that the parameter $b$ 
is model independent. We fix the subleading correction exponent $\epsilon=1.6$
or $\epsilon=2$. In the ansatz~(\ref{barfit3}) we take into account 
corrections $\propto L^{-3 \omega}$. Finally in the ansatz~(\ref{barfit3x})
we add, similar to eq.~(\ref{barfit2x}) a term $b L^{-\epsilon}$.

In the case of the ansatz~(\ref{barfit1}) fits with $\chi^2/$d.o.f.$<2$ are 
only obtained for $L_{min} \ge 36$.  
Instead, fitting with ansatz~(\ref{barfit2}) we get for $U_4$ at 
$Z_a/Z_p=0.5425$ $\chi^2/$d.o.f.$=62.4/62$ already for $L_{min}=16$.
The results for the parameters of this fit are
$\omega=0.832(1)$,  $\bar{U}_4^*=1.60357(1)$, 
$a(\mbox{Ising})=-0.2983(6)$, $a(0.641)=-0.0067(2)$, $a(0.655)=-0.0006(2)$,
$a(\ln 2)=0.0167(2)$, $a(1.15)=0.380(1)$, and $c=2.08(3)$. Note that here and 
in the following the errors quoted for results of individual fits are
purely statistical.
Extrapolating $a(0.641)$ and $a(0.655)$ we get $D^*=0.6564(5)$.

We estimated the systematic error due to corrections that are not taken 
into account in the ansatz~(\ref{barfit2}) from the variation of the 
results obtained with the 
ans\"atze~(\ref{barfit2x},\ref{barfit3},\ref{barfit3x}) and by using $U_6$
instead of $U_4$. Furthermore we have redone the analysis for $U_4$ and $U_6$
at $\xi_{2nd}/L=0.6431$. We arrive at the final estimates
\begin{eqnarray}
  \omega &=& 0.832(6) \\
    D^*  &=& 0.656(20) \;\;.
\end{eqnarray}
It also follows from the fits that the amplitude of corrections to scaling
at $D=0.655$ is at least by a factor of 30 smaller than that of the 
spin-1/2 Ising model. 

\section{Improved Observables} 
\label{improvedO}
The exponent $\nu$ can be obtained from the behavior of the slope 
of a phenomenological coupling at the critical point:
\begin{equation}
\label{slopeX}
\left . \frac{\partial R}{\partial \beta} \right |_{\beta=\beta_c} 
 = a L^{1/\nu} \;\;(1 + b L^{-\omega} + ...) \;\;.
\end{equation}
The exponent $\eta$ can be extracted from the behavior of the 
magnetic susceptibility at the critical point:
\begin{equation}
\label{chiX}
\left .\chi \right |_{\beta=\beta_c}
 = a L^{2-\eta} \;\;(1 + b L^{-\omega} + ...)  \;\;.
\end{equation}
Note that the coefficients $a$ and $b$ of course take different values in 
eq.~(\ref{slopeX}) and eq.~(\ref{chiX}).
Such a procedure requires an estimate of $\beta_c$.  To avoid this 
we have  studied, following \cite{Ha99}, the slopes and the magnetic 
susceptibility at $\beta_f$ as defined in eq.~(\ref{definebetaf}). 
These quantities behave as
\begin{equation}
\label{slopescaling}
\overline{\frac{\partial R}{\partial \beta}} \equiv \left . \frac{\partial R}{\partial \beta} \right |_{\beta=\beta_f}
 = a(D) L^{1/\nu} \;\;(1 + b(D) L^{-\omega} + ...)
\end{equation}
and 
\begin{equation}
\label{chiscaling}
 \bar{\chi} \equiv \left .\chi \right |_{\beta=\beta_f}
 = a(D) L^{2-\eta} \;\;(1 + b(D) L^{-\omega} + ...) \;\;.
\end{equation}
Again we have computed these quantities using their Taylor-expansion 
around $\beta_s$ up to the third order.

Here,  following \cite{ourdilute}, we shall study improved versions 
of the slopes and the magnetic susceptibility.  This means  in the ideal case
that the amplitude of leading corrections vanishes for any model.
In practice, as we shall see below, we can construct quantities for that
the amplitude of leading corrections is suppressed by more than one order
of magnitude.  Using such
quantities in the case of improved models ensures that leading corrections
to scaling are suppressed by two to three orders of magnitude compared with 
standard observables in the case of e.g. the spin-1/2 Ising model.
This is sufficient to ignore leading corrections to scaling in the analysis of 
our data.

Let us discuss in detail the construction of the improved observable 
at the example of the magnetic susceptibility. We consider
\begin{equation}
\label{defineI}
 \bar{\chi}_{imp}(L,D) = \bar{U}_4(L,D)^x  \bar{\chi}(L,D) \;\;,
\end{equation}
where $x$ is chosen such that the amplitude of leading corrections vanishes. 
Note that instead of $\bar{U}_4$ also $\bar{U}_6$ could be used.  It is 
important to take a phenomenological coupling, where leading corrections
to scaling are clearly visible.
Let us recall the finite size scaling behavior of the Binder cumulant
\begin{equation}
\label{baru}
 \bar{U}_4(L,D) = \bar{U}_4^* + b_U(D)  L^{-\omega} + ...  \;\;.
\end{equation}
Inserting   eq.~(\ref{chiscaling}) and eq.~(\ref{baru}) into 
eq.~(\ref{defineI}) we get
\begin{equation}
  \bar{\chi}_{imp}(L,D) =  a(D) \bar{U}_4^x  L^{2-\eta}
   \;\left(1 + \left[b(D) + x \frac{b_U(D)}{\bar{U}_4^*} \right] L^{-\omega} 
   + ...\right) \;\;.
\end{equation}
Hence the exponent defining the improved observable is given by
\begin{equation}
x = - b(D) \frac{\bar{U}_4^*}{b_U(D)}  \;\;.
\end{equation}
Note that ratios of correction amplitudes are universal. Therefore 
the exponent $x$ does not depend on $D$. It can be best determined 
by analyzing data obtained for models with relatively large corrections 
to scaling. For example one might consider the spin-1/2 Ising  model to 
this end.
We have already determined $b_U(\mbox{Ising})$ and $\bar{U_4}^*$ in the 
previous section. In order to obtain $b(D)$ one would fit 
$\bar{\chi}(L,D)$ with ans\"atze motivated by eq.~(\ref{chiscaling}).

However it turns out to be more efficient to study ratios of observables
taken from two different models. This way, critical exponents cancel and 
therefore fits have less parameters and become more reliable. 
In particular we shall study the  spin-1/2 Ising model and the Blume-Capel 
at $D=1.15$. 
We define 
\begin{equation}
 R_{\chi}(L) = \frac{\bar{\chi}(L,\mbox{Ising})}{\bar{\chi}(L,D=1.15)}
 = \frac{a(\mbox{Ising})}{a(D=1.15)} 
  \left(1 + [ b(\mbox{Ising}) - b(\mbox{1.15})] L^{-\omega} + ...
  \right)
\end{equation}
and 
\begin{equation}
 R_{U}(L) = \frac{\bar{U}(L,\mbox{Ising})}{\bar{U}(L,D=1.15)}
  = 1 + 
\frac{ b_U(\mbox{Ising}) - b_U(\mbox{1.15})}{\bar{U}^*} L^{-\omega} 
    + ... \;\;,
\end{equation}
where now
\begin{equation}
x = - [ b(\mbox{Ising}) - b(\mbox{1.15})] 
\frac{\bar{U}^*} { b_U(\mbox{Ising}) - b_U(\mbox{1.15})} \;\;.
\end{equation}
The exponent $x$ can be directly obtained from fits with the ansatz
\begin{equation}
\label{impro1}
  R_{U}(L)^x R_{\chi}(L) = C \;\;,
\end{equation}
where $x$ and $C$ are the parameters of the fit.
To check for the effect of subleading corrections we have also
fitted the data with the ansatz
\begin{equation}
\label{impro2}
   R_{U}(L)^x R_{\chi}(L)  = C + c L^{-\epsilon} \;\;,
\end{equation}
where $c$ is an additional parameter and 
we have fixed either $\epsilon=1.6$ or $\epsilon=2$.
Fixing $Z_a/Z_p=0.5425$, 
fits with the ansatz~(\ref{impro1}) have an $\chi^2/$d.o.f. $\approx 1$
starting with $L_{min} = 16$. Using $L_{min} = 16$ we get $x=-0.656(1)$. 
Instead using the ansatz~(\ref{impro2}) we get $\chi^2/$d.o.f. $\approx 1$ 
already for $L_{min}=10$. The results for $L_{min}=10$ are $x=-0.665(2)$ 
and $x=-0.661(2)$ for $\epsilon=1.6$ and $\epsilon=2$, respectively.
As our final result we quote $x=-0.66(1)$, where the error is chosen such 
that it covers the three estimates given above. In a similar fashion we
arrive at $x=-0.57(2)$ for fixing $\xi_{2nd}/L=0.6431$.

\begin{figure}
\begin{center}
\includegraphics[width=13.5cm]{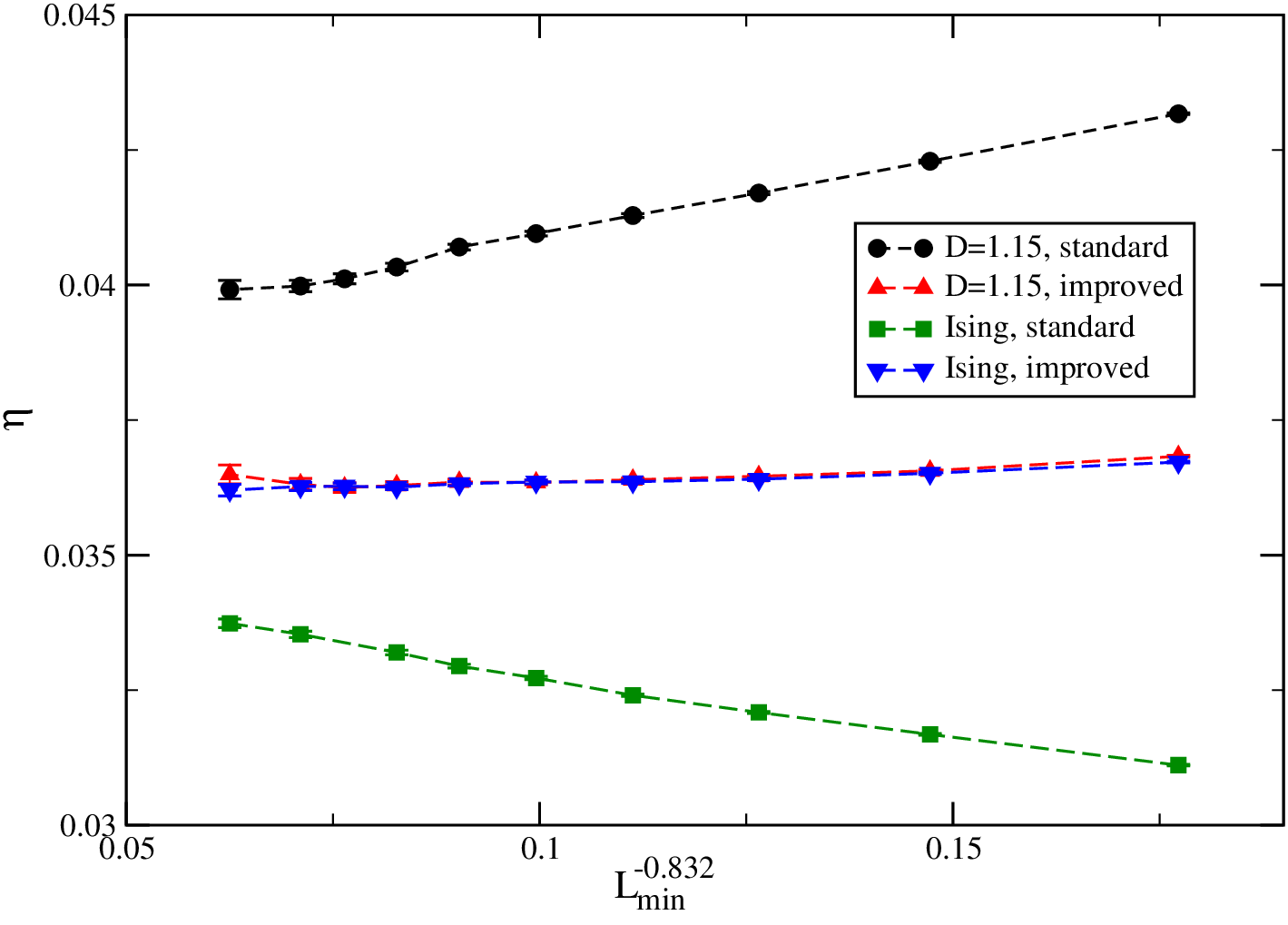}
\caption{\label{etademo}
Results for the critical exponent 
$\eta$ obtained by fitting the standard and the improved
magnetic susceptibility at $Z_a/Z_p=0.5425$ for the Ising model and the 
Blume-Capel model at $D=1.15$ using the ansatz~(\ref{chiback0}).
$L_{min}$ is the minimal lattice size that is taken into
account.
In the case of the improved magnetic susceptibility, the results obtained 
from the two different models fall nicely on top of each other.
The dashed lines should only guide the eye.
}
\end{center}
\end{figure}

In figure \ref{etademo} we demonstrate the effectiveness of the improvement. 
We have analyzed our data for $\chi$ at $Z_a/Z_p=0.5425$ for the Ising 
model and the Blume-Capel model at $D=1.15$. To this end, we have fitted
our data with the ansatz 
\begin{equation}
\label{chiback0}
\bar{\chi} = a L^{2-\eta} + B \;\;,
\end{equation}
where $B$ is an analytic background. 
Using the standard magnetic susceptibility, we get $\chi^2/$d.o.f.$=4.6/4$ 
for $L_{min}=32$ in the case of the Ising model and $\chi^2/$d.o.f.$=2.6/5$
for $L_{min}=24$ in the case of the Blume-Capel model at $D=1.15$. 
Nevertheless for e.g. $L_{min}=32$ the results for $\eta$ obtained from the 
two different models differ by more than 20 standard deviations.
In contrast, for the improved magnetic susceptibility the results obtained 
for the two models are quite similar. In particular for $L_{min}=24$ the 
estimates for $\eta$ obtained from the Ising model and the 
Blume-Capel model at $D=1.15$ are consistent within the error bars.

We also have constructed improved slopes 
\begin{equation}
\label{improveS1}
 \bar{S}_{imp}(L,D) =  \bar{U}_4(L,D)^x  \bar{S}(L,D) \;\;,
\end{equation}
where $x$ is chosen such that leading corrections to scaling vanish.
We have determined $x$ analogous to the case of the  magnetic susceptibility
discussed above. To this end we have computed the ratios  
\begin{equation}
 R_{S}(L) = \frac{\bar{S}(L,\mbox{Ising})}{\bar{S}(L,D=1.15)} \;\;.
\end{equation}
As discussed above for the case of the magnetic susceptibility, we have
fitted
\begin{equation}
R_{U}(L)^x R_{S}(L) = C
\end{equation}
with $x$ and $C$ as free parameters and, as check
\begin{equation}
R_{U}(L)^x R_{S}(L) = C + c L^{-\epsilon} \;\;,
\end{equation}
where $c$ is an additional parameter and $\epsilon$ is fixed to either 
$1.6$ or $2$. Our final results for the exponent $x$ are summarized in 
table \ref{XSlope1}.

\begin{table}
\caption{\sl \label{XSlope1}  Exponent $x$  of improved slopes as 
defined by eq.~(\ref{improveS1}). In the first column we give the
phenomenological coupling and its value that is used to define $\beta_f$.
In the first row we give the quantity whose slope  is considered.
For a discussion see the text.
}
\begin{center}
\begin{tabular}{ccccc}
\hline
fix ;  slope of   & $Z_a/Z_p$  &  $\xi_{2nd}/L$  & $U_4$ & $U_6$ \\
\hline
$Z_a/Z_p=0.5425$     & 0.52(2)  & 0.77(3) & -1.21(5) & -2.73(5)  \\
$\xi_{2nd}/L=0.6431$ & 0.54(2)  & 0.81(2) & -1.21(3) & -2.71(4)  \\
\hline
\end{tabular}
\end{center}
\end{table}

Furthermore  we have constructed quantities of the type 
\begin{equation}
\label{improveS2}
 \bar{S}_{ij} = |\bar{S}_i|^{x} |\bar{S}_j|^{1-x}  \;\;,
\end{equation}
where $x$ is again chosen such that the amplitude of the leading correction
vanishes. Here we performed fits with the ansatz
\begin{equation}
 R_{S_i}^x R_{S_j}^{1-x} = C \;\;,
\end{equation}
where $x$ and $C$ are the parameters of the fit and 
\begin{equation}
 R_{S_i}^x R_{S_j}^{1-x} = C + c L^{-\epsilon} 
\end{equation}
with the additional parameter $c$. Also here we have fixed either $\epsilon=1.6$
or $\epsilon=2$. In practice we have combined 
the slope of $U_4$ with the slope of $Z_a/Z_p$ or $\xi_{2nd}/L$. 
Our results for the exponent $x$ are summarized in table \ref{XSlope2}.
Notice that the results obtained for $Z_a/Z_p=0.5425$ and 
$\xi_{2nd}/L=0.6431$  are similar but not identical.

\begin{table}
\caption{\sl \label{XSlope2}   Exponent $x$ of improved  slopes as defined
by eq.~(\ref{improveS2}).
In the first column we give the
phenomenological coupling and its value that is used to define $\beta_f$.
In the first row we give the quantity whose slope  is mixed with that
of $\bar{U}_4$. For a discussion see the text.
}
\begin{center}
\begin{tabular}{ccc}
\hline
 fix ;  slope of  & $Z_a/Z_p$  &  $\xi_{2nd}/L$ \\
\hline
 $Z_a/Z_p=0.5425$ & 0.29(1) &  0.39(1)  \\
 $\xi_{2nd}/L=0.6431$  & 0.31(1) &  0.41(1)  \\
\hline
\end{tabular}
\end{center}
\end{table}

\section{The exponent $\eta$}
We have fitted our data for the improved magnetic susceptibility 
at $Z_a/Z_p=0.5425$ and $\xi_{2nd}/L=0.6431$ using the ans\"atze 
\begin{eqnarray}
\label{chisimple}
\bar{\chi}_{imp}  &=& a(D) L^{2-\eta} \\
\label{chiback}
\bar{\chi}_{imp}  &=& a(D) L^{2-\eta} + B(D) \\
\label{chieps}
\bar{\chi}_{imp}  &=& a(D) L^{2-\eta} \;\; \left(1 + d(D) L^{-\epsilon} \right)
\;\;. 
\end{eqnarray}
In the ansatz~(\ref{chiback}) we have taken into account the analytic 
background $B$ of the magnetic susceptibility. Since 
$\eta$ is small, the parameter $B$ also takes effectively into account 
other corrections that have a correction exponent $\omega'' \approx 2$
like for example the breaking of the rotational symmetry by the 
lattice.  In the ansatz~(\ref{chieps}) we have set $\epsilon=1.6$.  Using 
this ansatz we try to estimate the possible effect of a correction 
caused by $\omega'=1.67(11)$ \cite{NewmanRiedel} on our estimate of $\eta$. 

\begin{figure}
\begin{center}
\includegraphics[width=13.5cm]{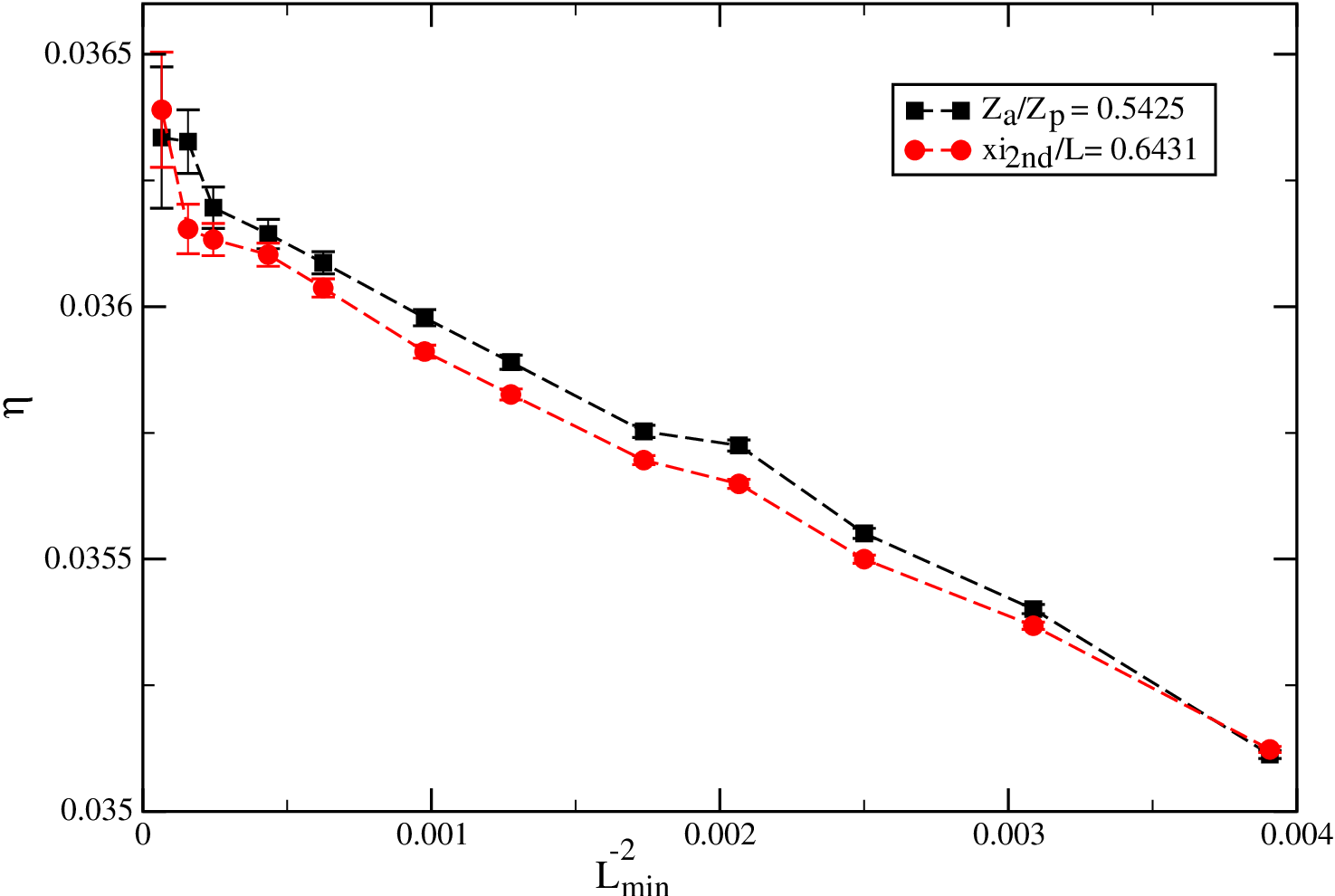}
\caption{\label{etafits}
Results for the critical exponent $\eta$ obtained  by fitting the 
improved magnetic
susceptibility at $Z_a/Z_p=0.5425$ and at $\xi_{2nd}/L=0.6431$ 
using the ansatz (\ref{chisimple}).  Data for the Blume-Capel 
model at $D=0.641$ and $D=0.655$ are taken into account.
$L_{min}$ is the minimal lattice size that is taken into 
account. 
The dashed lines should only guide the eye. For a discussion see the text.
}
\end{center}
\end{figure}

We have fitted the data for $D=0.641$ and $D=0.655$ in a common fit.
The parameters of these fits are $a(0.641)$, $a(0.655)$ and $\eta$ in the 
case of the ansatz~(\ref{chisimple}) and in addition 
$B(0.641)$ and $B(0.655)$ or $d(0.641)$ and $d(0.655)$ in the case of 
the ansatz~(\ref{chiback}) or (\ref{chieps}), respectively.
In figure \ref{etafits}  we have plotted  estimates of $\eta$  
obtained by fitting 
with the ansatz~(\ref{chisimple}) as a function of $L_{min}^{-2}$. 
Up to $L_{min} = 48$ the results fall roughly on a straight line, indicating
that corrections with an exponent $\epsilon \approx 2$ are present.
For $L_{min} = 128$ one finds that $\chi^2/$d.o.f. is smaller than one for 
both taking the improved susceptibility at $Z_a/Z_p=0.5425$ and 
$\xi_{2nd}/L=0.6431$.  The estimate $\eta=0.03636(20)$ covers both 
results obtained at $L_{min} = 128$, including their error bars.

\begin{figure}
\begin{center}
\includegraphics[width=13.5cm]{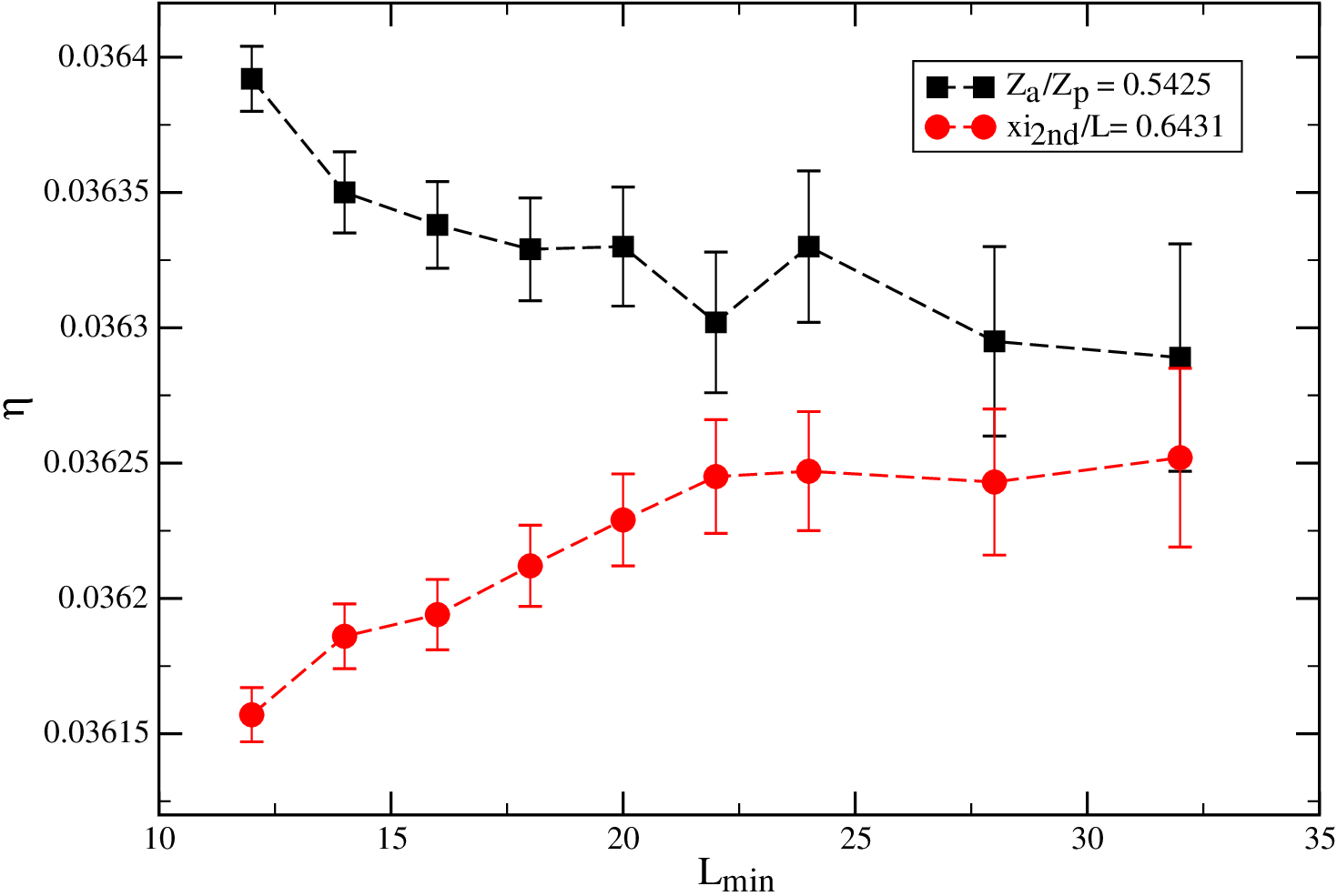}
\caption{\label{etafitsimp}
Results for the critical exponent $\eta$ obtained  by fitting the 
improved magnetic
susceptibility at $Z_a/Z_p=0.5425$ and at $\xi_{2nd}/L=0.6431$
using the ansatz (\ref{chiback}).  Data for the Blume-Capel
model at $D=0.641$ and $D=0.655$ are taken into account.
$L_{min}$ is the minimal lattice size that is taken into
account. The dashed lines should only guide the eye.
For a discussion see the text.
}
\end{center}
\end{figure}

Next we have fitted our data with the ansatz~(\ref{chiback}). For both
the improved magnetic susceptibility at $\xi_{2nd}=0.6431$ and at 
$Z_a/Z_p=0.5425$ we get $\chi^2/$d.o.f.$< 2$ starting from $L_{min}=14$.
The results for $\eta$ are plotted in figure \ref{etafitsimp}. In the 
case of $\xi_{2nd}=0.6431$ the estimate of $\eta$ is increasing with 
increasing $L_{min}$, while it is decreasing for $Z_a/Z_p=0.5425$. 
For $L_{min} =  32$ the two results are consistent within error bars. 

We read off our final estimate
\begin{equation}
\eta = 0.03627(10)  \;\;.
\end{equation}
The error estimate is chosen such that it also covers results 
obtained with the ansatz~(\ref{chieps})  and $L_{min}=32$. 

\section{The critical exponent $\nu$}
\begin{figure}
\begin{center}
\includegraphics[width=13.5cm]{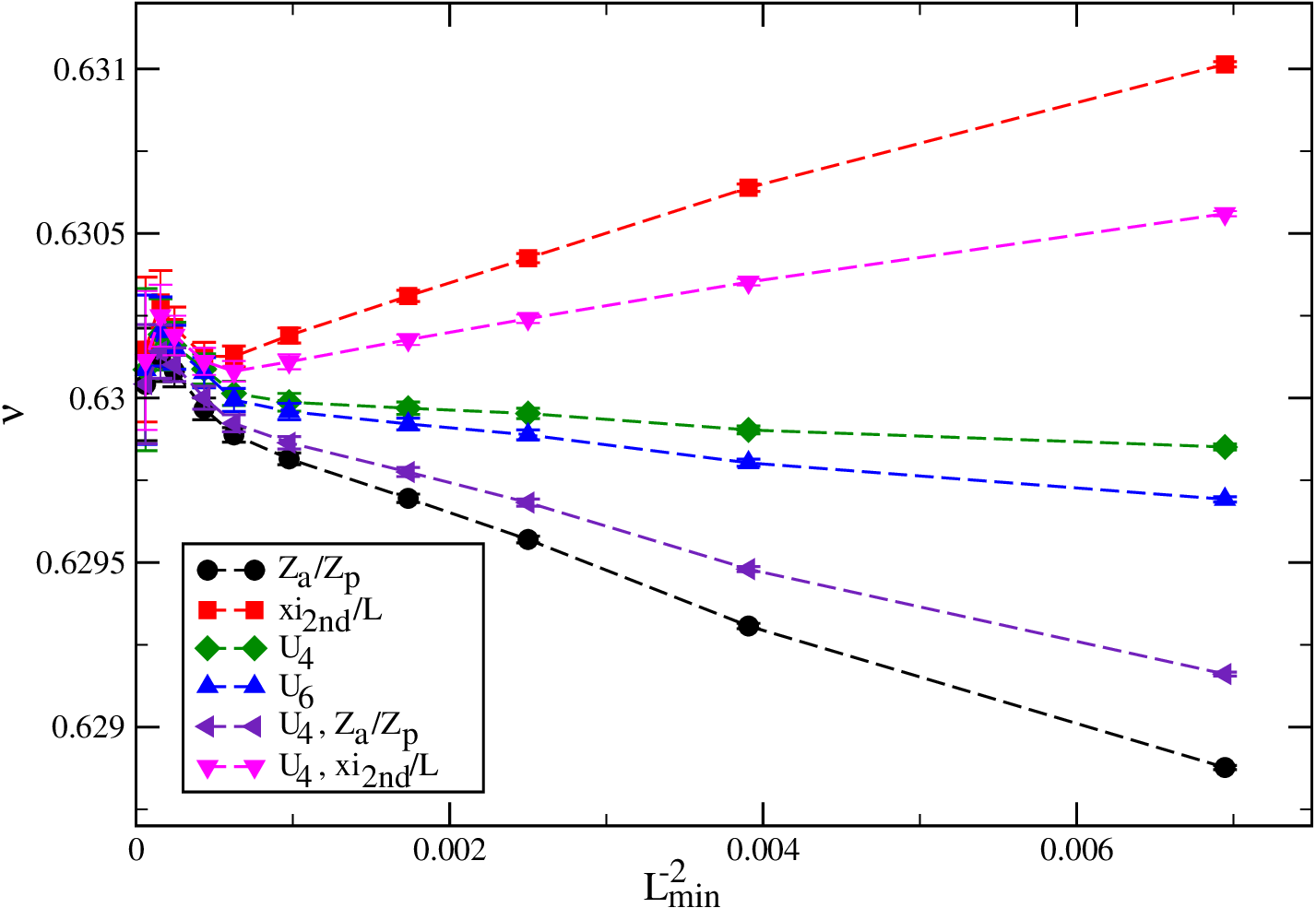}
\caption{\label{nuplot1}
Results for the critical exponent $\nu$ obtained by fitting improved 
slopes of various 
phenomenological couplings at $Z_a/Z_p=0.5425$ as a function 
of $L_{min}^{-2}$, where $L_{min}$ is the minimal lattice 
size that is included into the fit. The dashed lines should only guide the eye.
For a discussion see the text.
}
\end{center}
\end{figure}

In order to determine the exponent $\nu$ we performed combined fits 
of our data for the improved slopes at $D=0.641$ and $D=0.655$. 
In a first step of the analysis we have fitted the improved 
slopes with a power law without any correction
\begin{equation}
 S = a(D) L^{1/\nu} \;\;,
\end{equation}
where the amplitudes $a(0.641)$, $a(0.655)$ and the exponent $\nu$ are the 
parameters of the fit. In figure \ref{nuplot1} we give the results for $\nu$ 
as a function of $L_{min}^{-2}$, where $L_{min}$ is the minimal 
lattice size that is included into the fit. In the figure we give only 
results for taking the slopes at $Z_a/Z_p=0.5425$. 
Those for the slopes at $\xi_{2nd}/L=0.6431$ behave in a very similar way.

We find that the result for $\nu$ obtained from the improved slope of 
$Z_a/Z_p$ is increasing with increasing $L_{min}$, while the one obtained from
$\xi_{2nd}/L$ is decreasing.  The results obtained from the improved 
slope of $U_4$ and $U_6$ are quite similar. They only slightly increase 
with increasing $L_{min}$. 
We have also plotted results obtained from the combined 
slopes~(\ref{improveS2}). In the case of combining the slope of $U_4$ with
that of $\xi_{2nd}/L$ the estimate of $\nu$ is decreasing with increasing 
$L_{min}$, while for combining the slope of $U_4$ with that of 
$Z_a/Z_p$ it is increasing. In all cases a rather large $L_{min}$ is need to 
get acceptable values for $\chi^2/$d.o.f. . In the worst case, 
for the improved slope of
$Z_a/Z_p$  only for $L_{min} \ge 56 $ a $\chi^2/$d.o.f. smaller than two
is reached. The behavior of the estimates of $\nu$ for $L_{min} < 48$ 
is consistent with the fact that the dominating corrections have an 
exponent $\omega' \approx 2$. For larger $L_{min}$, the variation of 
our estimates of $\nu$ with $L_{min}$ seems to be dominated by 
statistical fluctuations. For $L_{min}=128$ we get $\chi^2/$d.o.f. smaller 
than one for all quantities that we have considered. The estimate 
$\nu=0.6301(3)$ covers the results, including the statistical error, 
of all our fits for $L_{min}=128$. Note that in ref. \cite{DengBloete}  
$L=128$ is the largest lattice size that is simulated.

\begin{figure}
\begin{center}
\includegraphics[width=13.5cm]{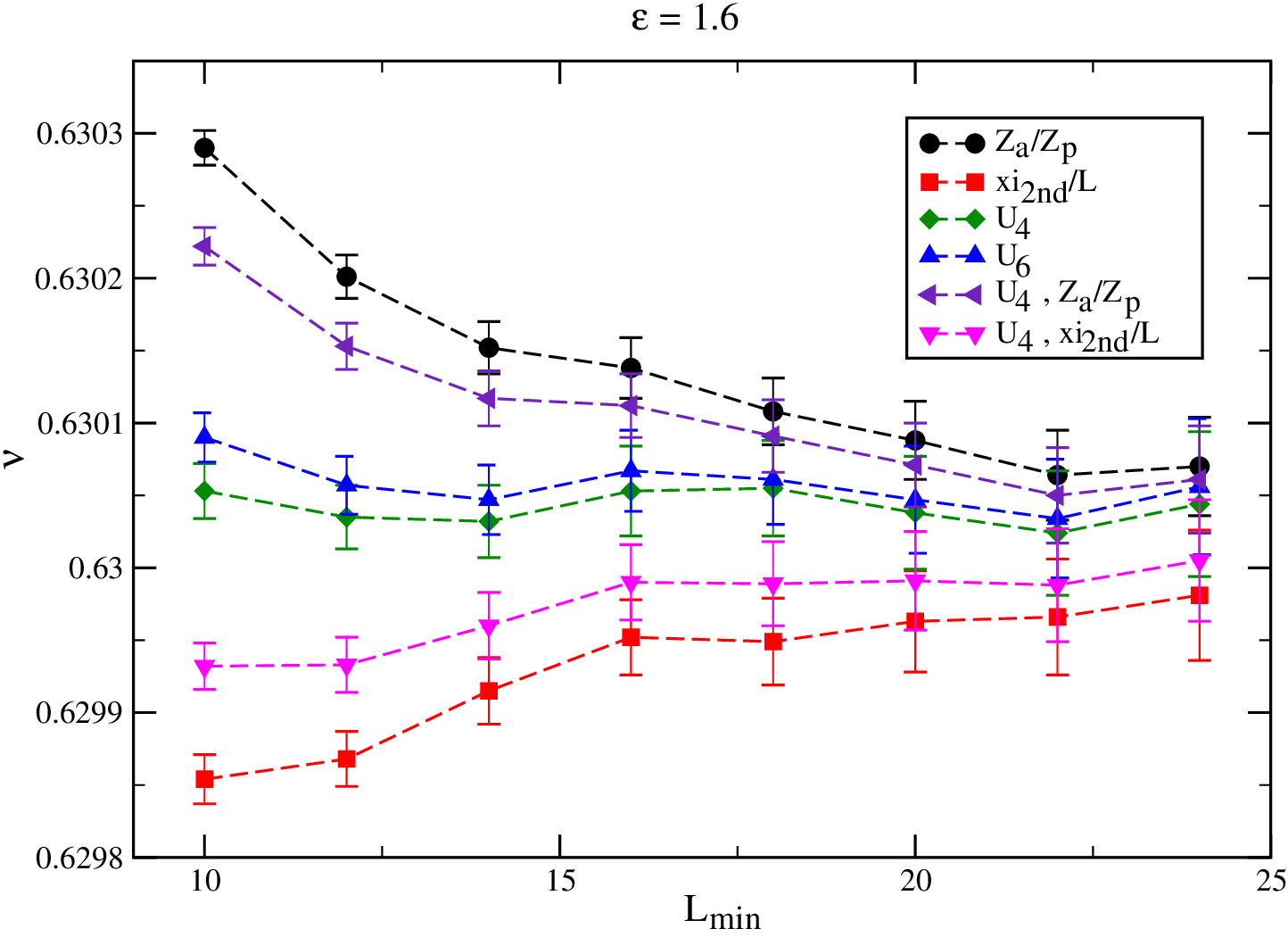}
\vskip1.1cm
\includegraphics[width=13.5cm]{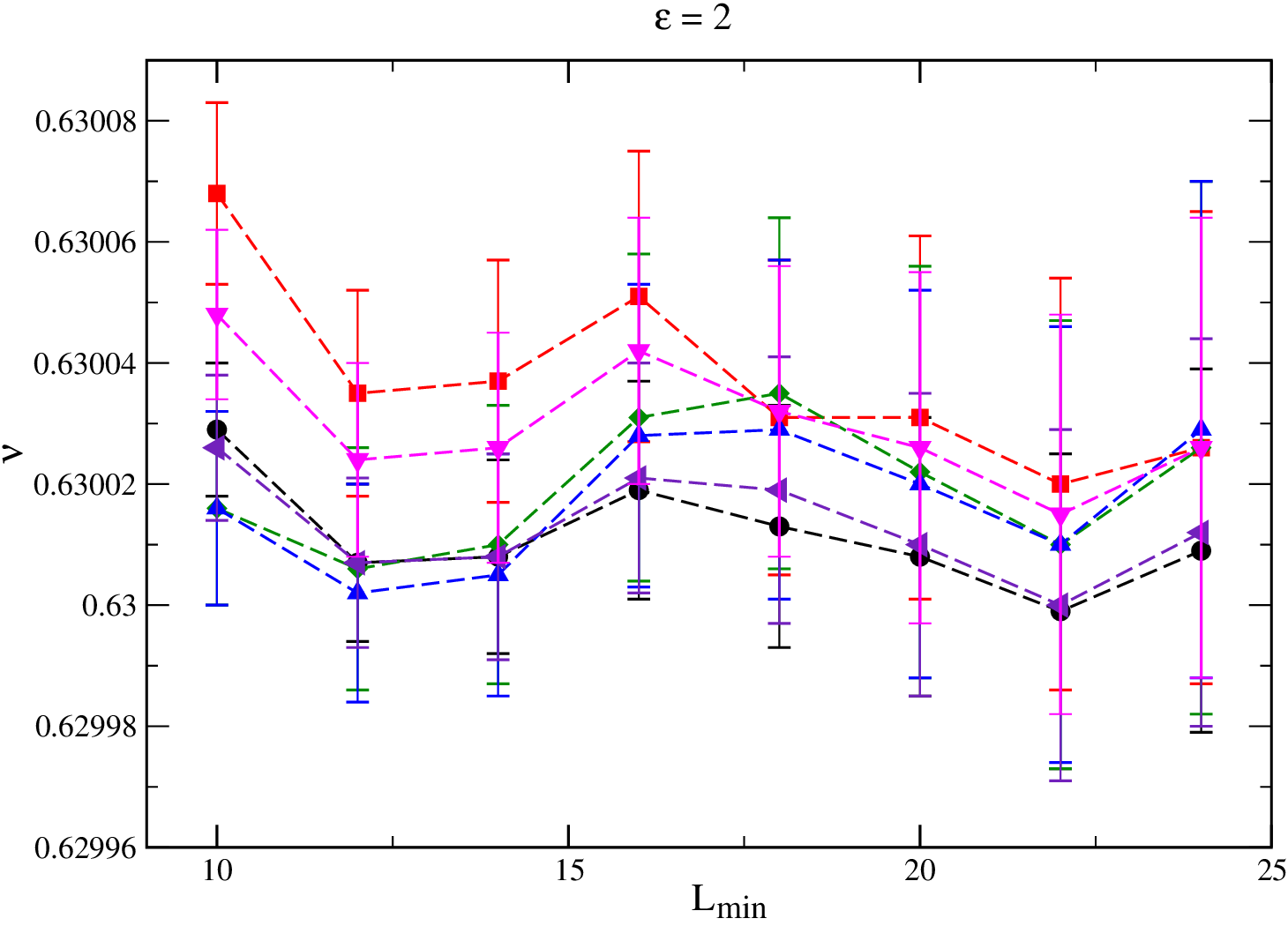}
\caption{\label{nuplot2}
Results for the critical exponent $\nu$ obtained by fitting 
improved slopes of various
phenomenological couplings at $Z_a/Z_p=0.5425$ with the 
ansatz~(\ref{nucorrection}) as a function of $L_{min}$. In the upper
part of the figure the correction exponent is fixed to $\epsilon=1.6$
and in the lower part it is fixed to  $\epsilon=2$.
The dashed lines should only guide the eye.
For a discussion see the text.
}
\end{center}
\end{figure}

Motivated by these observations, we have fitted our data with the ansatz
\begin{equation}
\label{nucorrection}
 S = a(D) L^{1/\nu} \times (1 + b L^{-\epsilon}) \;\;,
\end{equation}
where we have fixed  $\epsilon$ to either $1.6$ or $2$.  Since 
already $a(0.641)$ and $a(0.655)$ are very similar, we have chosen 
the parameter $b$ to be model independent.  Let us first discuss 
the fits with  $\epsilon=2$.  Such fits give $\chi^2/$d.o.f. close to 
one already for $L_{min}=10$.  In the lower part of figure \ref{nuplot2} 
we have plotted the results obtained from the slopes of different quantities
for $10 \le L_{min} \le 24$.
These different estimates of $\nu$ are consistent among each other.
Furthermore there is little variation of the results with $L_{min}$.

In the upper part of figure \ref{nuplot2} we plot the corresponding 
result for $\epsilon=1.6$.  
Here the $\chi^2/$d.o.f. is somewhat larger than for $\epsilon=2$. 
Also the result for $\nu$ clearly depends on the quantity that is 
analyzed.  We conclude that the numerically dominant corrections 
have an exponent $\epsilon \approx 2$. Motivated by these fits, 
we take $\nu=0.63002$ as our final result. Since we can not exclude
that there are also corrections with an exponent $\epsilon \approx 1.6$, 
we take these fits into account in our final error of $\nu$. 
For $L_{min}=22$ and 
$L_{min}=24$ all results that we have obtained with the 
ansatz~(\ref{nucorrection}), including their error
bar, are contained in the interval $[0.62992,0.63012]$. Therefore we 
quote as our final result
\begin{equation}
\nu=0.63002(10) \;\;.
\end{equation}

\section{Summary and Conclusions}
We have simulated the spin-1/2 Ising model and the Blume-Capel model 
on the simple cubic lattice using linear lattice sizes $L \le 360$.
Using finite size scaling methods we have determined critical properties
of these models. In particular we have determined the value
$D^*=0.656(20)$  of the parameter $D$ of the Blume-Capel model,
where leading corrections to scaling vanish. We have
accurately determined the inverse of the critical temperature for various 
values of $D$, in particular $\beta_c(0.641)=0.38567122(5)$ and 
$\beta_c(0.655)=0.387721735(25)$. We have computed the critical exponents
$\nu=0.63002(10)$ and $\eta=0.03627(10)$ as well as the exponent 
$\omega=0.832(6)$ of leading corrections to scaling. The errors 
quoted for these final results cover statistical as well as systematical
errors. Systematical errors are due to the fact that power laws like
eqs.~(\ref{slopeX},\ref{chiX}) that govern the finite size scaling 
behavior of physical quantities at the critical temperature
are subject to an infinite series of correction terms. Fitting Monte Carlo
data, only few of these correction terms can be taken into account. In the 
present study, we have effectively eliminated the leading correction 
$\propto L^{-\omega}$ by simulating an improved model and analyzing 
improved observables as discussed in section \ref{improvedO}. In our 
ans\"atze we take into account a sub-leading correction with the exponent
$\omega' =1.67(11)$ predicted by \cite{NewmanRiedel} or $\omega''\approx 2$
due to the breaking of the rotational symmetry by the simple cubic 
lattice \cite{pisa97}
or due to the analytic background of the magnetic susceptibility. We 
estimate the error caused by correction terms that are not included 
by comparing the results obtained by using different ans\"atze and, even 
more important, by fitting different quantities. One expects that in the 
generic case the amplitudes of corrections are different for different 
quantities. In the case of the critical exponent $\nu$ we have studied the 
slope of four different phenomenological couplings: The cumulants $U_4$ and
$U_6$, the ratio of partition functions $Z_a/Z_p$, and the second moment 
correlation length over the linear lattice size $\xi_{2nd}/L$.  We regard
the estimates of the error obtained this way as quite robust and therefore
the results obtained
here should serve well as benchmark for experimental studies as well as
new or developing theoretical methods.

Our results are fully consistent with those obtained from high temperature
series expansion of lattice models \cite{pisa_series2,BuCo02,BuCo05}; See 
table \ref{compare_lattice}.
We find a small discrepancy with the Monte Carlo results of ref. 
\cite{DengBloete}; See table \ref{compare_lattice}. 
Note that the authors of  \cite{DengBloete}  did
not take into account a sub-leading correction with the exponent 
$\omega' =1.67(11)$ \cite{NewmanRiedel} analyzing their Monte Carlo data.

The accuracy that is reached now 
by lattice methods has clearly outpaced that of field theoretic 
methods. Furthermore, comparing with the numbers that are summarised in
table \ref{compare_field}, we notice that most of the results for 
$\eta$ and $\omega$ obtained from the perturbative expansion in three 
dimensions fixed are at odds with ours, while those of \cite{Guelph, Nickel}
are in reasonable agreement. Note that, as discussed by
Nickel \cite{Nickel}, the subleading correction exponent 
$\omega' =1.67(11)$ \cite{NewmanRiedel} also  plays a crucial role 
in the analysis of the perturbative series in three dimensions fixed. 
Therefore, it would be highly desirable to get an estimate of $\omega'$
by using a different method.

Using Monte Carlo simulations, the error of the estimates of the 
critical exponents can be further reduced just by spending more 
CPU time. To this end one has to increase  the statistics as well as 
enlarge the size of the lattices that are simulated.  
Keeping the statistical error and the systematical one proportional, the 
effort increases as $\mbox{error}^{-2 - (3+z)/\omega'}$ with a decreasing
error, where the first factor $\mbox{error}^{-2}$ is related to the increased
statistics and the second to the larger linear lattice size $L$ that is needed 
to reduce the systematical error. Here we assume that the systematical error
is proportional to $L^{-\omega'}$, since, as we have shown here, 
leading corrections can be eliminated. The effort at a fixed statistical 
accuracy behaves as $L^{d+z}$, where $d=3$
is the dimension of the system and $z$ is the critical dynamical exponent.
In a recent study of a spin glass \cite{glass} about 1000 years of 
CPU time on one core of a CPU of similar performance as the one used 
here had been spent. This is about a factor of 30 more CPU time than we have
spent here. One should notice however that this factor in CPU time only 
would allow to reduce the errors of the critical exponents by a factor of 
about $2.3$, where we have assumed $\omega' \approx 1.6$ and $z \approx 0.4$; 
see Eqs.~(\ref{tau1},\ref{tau2}).
\section{Acknowledgements}
This work was supported by the DFG under the grant No HA 3150/2-1.

\end{document}